\documentclass[fleqn,10pt]{article}
\usepackage{amsmath}
\begin{document}


\title{Unitary and Lorentz invariance in QCD for various gauges}

    \author{A. Andra\v{s}i,\\
Vla\v{s}ka 58, Zagreb, Croatia.(andjelka.andrasi@irb.hr) \\J. C. Taylor,\\
DAMTP, Cambridge University, UK (jct@
cam.ac.uk)}
\maketitle

\begin{abstract}
We generalise the Kugo and Ojima formalism (henceforth called KO), for the structure of the state-space in gauge theories, in four respects:- (i) We allow for a more general class of gauge-fixing, including non-covariant cases. (ii) We display the two possible Hamiltonians allowed by the KO action. (iii) We give manifestly Lorentz covariant forms of KO, showing that they are not unique. (iv) We analyse the structure of Fock space, proving that the allowed states span exactly half of it (leaving aside the pure transverse states).
\pagebreak
\end{abstract}

\maketitle
\

\section*{Introduction}
The structure of the state-space in gauge theories is well understood, since at least the classic paper of Kugo and Ojima \cite{key1} in 1987 (henceforth referred to as KO). They introduce an auxiliary field $B^a(x)$ in addition to the gauge field $A_\mu^a(x)$ ($a$ is the colour index); and hence define a conserved, nilpotent BRST operator  $Q$. Physical states are represented
by
\begin{equation}
\rm{im}Q/\ker Q 
\end{equation}
that is to say state vectors $|X>$ satisfying
\begin{equation}
    Q|X>=0
\end{equation}
with
\begin{equation}
    |X'>=|X>+Q|Y>
\end{equation}
representing the same physical state for any $|Y>$. These state-vectors all have non-negative norm. This is summarized in most modern textbooks, for example \cite{key2} section 15.7.
    KO deal with a general covariant gauge, with a parameter $\alpha$, in which the gluon propagator is 
\begin{equation}
  \frac{[g_{\mu\nu} -(1-\alpha^2)k_{\mu}k_{\nu}/ k^2]}{k^2}.
\end{equation}
This gauge is treated in many textbooks, and Feynman perturbation theory appears to work for it, in spite of the double pole. But when we apply the KO formalism to it, serious complications result, because the field equations are fourth order in time derivatives.
One purpose of the present paper is to draw attention
to these complications, which are somewhat hidden in KO.
     We also generalise the class of gauges to include
non-covariant gauges, by introducing another parameter, $\theta$. This allows a limit to the Coulomb gauge. The important parameter is then
$\frac{\alpha^2}{\theta^2}-1$ which we call $s$. Problems arise unless $s=0$.

The plan of the paper is as follows. In section 1, we state the generalized KO action which we use, and the field equations. 
In section 2, we derive the two possible Hamiltonians (related by a canonical transformation).
We note that the free Hamiltonian is not diagonal unless $s=0$. In section 3, we state the expansions of the asymptotic fields in terms
of creation and annihilation operators. We also give the Fourier transform of the field.  The four-momentum is not in general lightlike unless $s=0$. In section 4, we deduce the form of $Q$ in both Hamiltonian formalisms.
Section 5 contains a manifestly Lorentz covariant version of the KO theory, which is also shown to be non-unique.
Section 6 stresses the importance of pure transverse states.
Section 7 is an analysis of the structure of the complete Fock space. We show that allowed states span exactly half the Fock space, leaving aside pure transverse states.

\section{The KO Lagrangian}
The Lagrangian we will use is a slight generalisation of that used by KO, allowing for non-covariant gauges (we restrict to pure Yang-Mills theory - it makes no important difference to include quarks). The Lagrangian is
\begin{equation}
    L=\int d^3x (\mathcal{L}(x)+\mathcal{L}
    _G (x))
\end{equation}
where
\begin{equation}
      \mathcal{L}=-\frac{1}{4}F_{\mu \nu}F^{\mu \nu}-\theta^2 A_0\partial_0 B+B\partial_i A_i+\frac{1}{2}\alpha^2 B^2 
    \end{equation}
    \begin{equation}
\mathcal{L}_G=\partial_i\hat{c}D^i c+\theta^2\partial_0\hat{c}D^0c.
\end{equation}
Here $c$ is the ghost field and $\hat{c}$ the anti-ghost. Colour indices are not shown explicitly. KO's
Lagrangian is got by setting $\theta=1$. The Coulomb gauge is obtained in the limit
$\theta^2=\alpha^2 \rightarrow 0$.

There is an alternative Lagrangian, which we will call $L'$, got  from (6) by replacing  $-A_0\partial_0 B$ by $B\partial_0 A_0$. This gives the same action and field equations as $L$, but leads to a different Hamiltonian formulation. 

To write the gluon propagator, we introduce the notation, for a 4-vector $k_\mu$, (we denote spatial indices by Latin letters)
\begin{equation}
    K^2=k_ik_i, \ k^2=k_0^2-K^2, \ k_\theta^2=\theta^2k_0^2-K^2.
\end{equation}

The field equations are
\begin{equation}
      D^\mu F_{\mu  i}=\partial_i 
 B+g\partial_i\hat{c}\wedge c,
\end{equation}
\begin{equation}
    D^i(D_i A_0 -\dot{A}_i)=\theta^2(\dot{B}+g\partial_0 \hat{c}\wedge c),
\end{equation}
\begin{equation}
 \alpha^2 B=-\theta^2\partial_0 A_0-\partial_i A^i,  
\end{equation}
where $D^{\mu}$ is the covariant derivative.

     Equation (10) does not contain a time derivative of $A_0$ and (11) does not contain a time derivative of $B$. So there are two options. (11) could be used to eliminate $B$, obtaining the ordinary Yang-Mills Lagrangian. But $B$ seems to be essential
to the KO argument, so they do not do this.  In principle, (10) could be used to eliminate $A_0$, leading to a non-local effective Lagrangian (not explicitly Lorentz invariant, even for $\theta =1)$.  KO take neither of these options, preferring to work with the original Lagrangian (6), although there are the constraints (10) or (11). We will show below that, in the Hamiltonian formalism, the above two options are equivalent, up to a trivial canonical transformation. So there is actually no need to introduce the $B$ field, if the Hamiltonian formalism is used.

    For the free, asymptotic fields, we deduce from (9), (10) and (11) that
\begin{equation}
   (\theta^2 \partial_0^2-\partial_i\partial_i )B=0.
\end{equation}
The components of $A_\mu$ do not all obey similar homogeneous wave equations. For instance
\begin{equation}
  (\theta^2\partial_0^2-\partial_i\partial_i)A_0 =(\theta^2-\alpha^2)\partial_0 B,
\end{equation}
\begin{equation}
  (\theta^2\partial_0^2-\partial_i\partial_i)A_j^L=(\theta^2-\alpha^2)\partial_j B,
\end{equation}
\begin{equation}
  (\partial_0^2-\partial_i\partial_i)A_j^T=0,
\end{equation}
where we write
\begin{equation}
  A_j=A_j^L+A_j^T   
\end{equation}
with
\begin{equation}
\partial_jA_j^T=0.
    \end{equation}
This Lagrangian is invariant under the following increments:
\begin{equation}
    \delta A_\mu=\eta D_\mu c,\  \delta c=-\eta gc\wedge c/2,\  \delta \hat{c}= \eta B, \ \delta B=0.
\end{equation}
where $\eta$ is a Grassmann number. (Nilptency can be expressed as $\delta' \delta =0$,
where $\delta'$ is defined like $\delta$ but with a different Grassmann  number $\eta'$.)

\section{The KO Hamiltonians}

Let $\Pi_i$ be the momentum conjugate field to  $A_i$,  ($i=1,2,3$) and $\Pi_B$  be conjugate to $B$. Then
\begin{equation}
    \Pi_i=F_{0i}=\dot{A}_i-D_iA_0
\end{equation}
($D$ is the covariant derivative), and
\begin{equation}
    \Pi_B=-\theta^2A_0,
\end{equation}
\begin{equation}
    \Pi_c=\theta^2\partial_0\hat{c},  \ \Pi_{\hat{c}}=-\theta^2D_0c.
\end{equation}Using  the above equations, we find that 
\begin{equation}
    H+H_G=\int d^3\mathbf{x}(\mathcal{H}+\mathcal{H}_G),
\end{equation}
\begin{equation}
\mathcal{H}=\frac{1}{2}\Pi_i(\Pi_i-2\theta^{-2}D_i \Pi_B) +B\partial_iA_i +
\frac{1}{4}F_{ij}F_{ij}-\frac{1}{2}\alpha^2B^2 
\end{equation}
\begin{equation}
\mathcal{H}_G =\theta^{-2}\Pi_{\hat{c}}
\Pi_c+\partial_i\hat{c}D_i c+\theta^{-2}g\Pi_c.(\Pi_B\wedge c).
\end{equation}
(For fermionic fields, we define $H$ with the ordering $p\dot{q}-L$, so we have used right-differentiation to get $\Pi_c$ and $\Pi_{\hat{c}}$.)

    The equal-time commutation relations are
\begin{equation}
[A_i^a(\mathbf{x},t),\Pi_j^b(\mathbf{y},t)]=i\delta^3(\mathbf{x}-\mathbf{y})\delta_{ij}\delta^{ab},
    \end{equation}
    \begin{equation}
[A_0^a(\mathbf{x},t),B^b(\mathbf{y},t)]=i\theta^{-2}\delta^3(\mathbf{x}-\mathbf{y})\delta^{ab}
    \end{equation}

where we have made explicit the  colour indices $a,b$. Also the anticommutators

\begin{equation}
\{c({\bf{x}},t),\dot{\hat{c}}({\bf{y}},t)\}=i\theta^{-2}\delta({\bf{x}-\bf{y})},
\end{equation}
\begin{equation}
 \{\hat{c}({\bf{x}},t),\dot{c}({\bf{y}},t)\}=-i\theta^{-2}\delta({\bf{x}-\bf{y}}).   
\end{equation}
Note that $\hat{c}$ is antihermitean.

          If we choose the option to use the Lagrangian $L'$ , defined after (6), we get the Hamiltonian
\begin{equation}
\mathcal{H}'=\frac{1}{2}\Pi_i(\Pi_i+2D_i A_0)+\theta^{-2}\Pi_0 \partial_i A_i +
\frac{1}{4}F_{ij}F_{ij}-\frac{\alpha^2}{2\theta^4}\Pi_0^2
\end{equation}
This results in the use of the constraint (11) to eliminate the KO auxiliary field $B$.
The ghost part of $H'$ is the same as for $H$   in (24) .
This ${H'}$ is related to ${H}$ in (22) by the canonical transformation
\begin{equation}
    B \rightarrow \theta^{-2}\Pi_0, \ \Pi_B \rightarrow -\theta^{2}A_0,
\end{equation}

and so has all the same properties. The commutation relation becomes  
\begin{equation}
        [A_0^a(\mathbf{x},t),
        \Pi_0^b(\mathbf{y},t]=i\delta^3(\mathbf{x}-\mathbf{y})\delta^{ab}.
    \end{equation}
    In this Hamiltonian approach, the role of $B$ is taken over by $\Pi_0$. 
    
\section{The free fields}

We take the 4-momenta of the transverse gluons to be $(P;\bf{p})$ and of the time-like and longitudinal gluons to be $(P_\theta;\bf{p})$, where we define $P\equiv |\bf{p}|$ and $P_\theta \equiv P/\theta$.

We define polarization 4-vectors as follows.
\begin{equation}
e_T^m=(0;\mathbf{e}_T^m) \ (m=1,2), \
e_\pm=(\mp 1;{\bf{p}}/ P_\theta)/\sqrt{2},
\end{equation}
\begin{equation}
    \mathbf{p}.\mathbf{e}_T^m=0,\ \mathbf{e}_T^m.\mathbf{e}_T^n=\delta^{mn},
\end{equation}
thus, for $\theta=1$, $e_-$ is parallel to the (lightlike) 4-momentum of the gluon.
These are in the particular coordinate system, in which the transverse quanta are purely spacelike. 

We introduce annihilation operators, $a_T^{(m)},a_\pm$, satisfying
\begin{equation}
    [a_T^{(m)}({\bf{k}}),{a_T^{(n)}}^*({\bf{k'}})]=\delta^{mn}\delta({\bf{k}-\bf{k'}}),\,[a_+({\bf{k}}),a_-^*({\bf{k'}})]=\delta({\bf{k}-\bf{k'}})
\end{equation}
(colour indices being omitted). All other commutators are zero.
In addition, there are ghost operators $u,v$, obeying
\begin{equation}
    \{u({\bf{k}}),v^*({\bf{k'}})\}=\delta({\bf{k}}-{\bf{k'}}).
\end{equation}
    With this notation, the expression given by KO for $A_\mu$ , generalised to the gauge (6), may be written
    \begin{equation*}
      A_{\mu}(x)=  \frac{1}{\theta}\int_{\theta}[(1+\frac{1}{2}s)e_{+\mu}a_- +e_{-\mu}a_+ +(istP_\theta)e_{-\mu}a_-]E_\theta
    \end{equation*}
    \begin{equation}
        +\sum_m\int e_{T\mu}^{(m)} a_T^{(m)}E +\rm{hermitean}\, \rm{conjugates}.
    \end{equation}
    Here we define
\begin{equation}
        \int_\theta\equiv \int d^3{\bf{p}}
((2\pi)^3 2P/\theta)^{-1/2 }, \ \, E_\theta \equiv \exp{(i{\bf{p}}.{\bf{x}}-iP_{\theta} t)}  
\end{equation}
and $\int$ and $E$ are defined similarly with $\theta$
replaced by 1. It is understood that everything in the integrands in ( 36) are functions of $\mathbf{p}$.
Also
\begin{equation}
    s\equiv  \frac{\alpha^2}{\theta^2}-1.
\end{equation}
(Equation (36), which seems to us a key equation in the KO method, does not appear explicitly in this form in [1].)

Similarly
\begin{equation}
  B(x)=-i\frac{\sqrt{2}}{\theta}\int_{\theta} P_\theta [a_- E_{\theta}  -a^*_- E^*_{\theta}],
\end{equation}
and the conjugate momentum to $A_i$ is
\begin{equation}
\Pi_i=-i\frac{\sqrt{2}}{\theta}\int_\theta p_i a_- E_\theta
 -i\sum_m\int P a_T^m e_T^m E \ +(\mathrm{c.c.})
\end{equation}
It can be verified that (36) and (39), satisfy the equations of motion
(9), (10), (11) and the commutation relations (25) and ( 26).  (These conditions do not specify $A_\mu$ uniquely
however, as shown in section 5.)

The noteworthy point about (36) is the explicit dependence on the time $t$ (for $s \not=0$).
This comes from a solution of equations (13 ) and (14). On the face of it, this
might break  Lorentz invariance. Although the field equations are covariant (for $\theta=1$), the equal-time commutation relations are not.

The free ghost fields have the forms
\begin{equation}
    c(x)=-(1/\sqrt{2})\int_\theta P^{-1}(u({\bf{k}})E+u^*({\bf{k}})E^*),
    \end{equation}
    \begin{equation}
    \hat{c}=(\sqrt{2}/\theta^2)\int_\theta P(vE-v^* E^*),
\end{equation}
where $u,v$ obey the anticommutation relation (35).

The equation  (36) for $A_\mu(x)$ can be cast into another form. We give only the special case $\theta=1$, and write it in two parts, the part proportional to $s$ and the rest:
\begin{equation}
  \frac{s}{\sqrt{2}}\int d^4k\gamma e^{-ik.x}k_\mu [\delta'(k_0-K)+(1/2K)\delta(k_0-K)]a_-,
\end{equation}
\begin{equation}
\int d^4k \gamma e^{-ik.x}\delta(k_0-K)[(1/\sqrt{2}K)k_\mu a_+ 
 +e_{+\mu}a_-)+\sum e^m_\mu a^m],
\end{equation}
where $\gamma =[2K(2\pi)^ 3]^{-1/2}$.
The factor $k_\mu$ in (43) and (44), allows the use of generalized Ward identities. We note that
we cannot use $k^2=0$ in general because of the derivative of the delta function.

For asymptotic fields, the Hamiltonian $H_0$ is
\begin{equation}
 \int d^3\mathbf{p}[P(\sum_m a_T^{(m)*}a_T^{(m)}) +P_\theta(a_+^* a_- + a_-^* a_+-sa^*_- a_- +u^* v+v^* u)].
\end{equation}
For $s$ nonzero, no asymptotic states containing  any $a_+^*$ quanta can be eigenstates of $H_0$.

Also for $s$ nonzero, $H_0$ together with the momentum operator do not constitute a
Lorentz 4-vector.

Nevertheless, from the above equations we can calculate the propagators and the unequal time commutation relations (of the free fields), and they prove to be covariant for $\theta=1$. They are consistent with the free field equations, of course.

For a 4-vector $k_\mu$, we use the notation
\begin{equation}
    K^2=k_ik_i, \ k^2=k_0^2-K^2, \ k_\theta^2=\theta^2k_0^2-K^2.
\end{equation}
Defining $D_\mu (k)$ by
\begin{equation}
    <0|T(B(x)A_\mu (y))|0>=-i(2\pi)^{-4}\int d^4k D_\mu (k) e^{-ik.(x-y)}
\end{equation}

\begin{equation}
    D_\mu =ik_\mu(k_\theta^2)^{-1}.
\end{equation}

Defining the $(A_\mu A_\nu)$ propagator similarly as
\begin{equation}
    D_{\mu\nu}+\hat{D}_{\mu\nu}
\end{equation}
where,
\begin{equation}
    D_{00}=(k_\theta^2)^{-1}, \ D_{0i}=0,\ D_{ij}=-\delta_{ij}(k^2)^{-1}+(k_ik_j/K^2)((k^2)^{-1}-\theta^2(k_\theta^2)^{-1})
\end{equation}
and
\begin{equation}
    \hat{D}_{\mu\nu}=(\alpha^2 -\theta^2)k_\mu k_\nu (k_\theta^2)^{-2}.
\end{equation}

 In all the above equations, it is to be understood that the denominators (including the double denominators) have an infintesimal imaginary part
 $+i\eta$.
 
The unequal time commutators of the free
gluon field can be obtained from the propagators by the substitution
\begin{equation}
    (k_\theta^2+i\eta)^{-1} \rightarrow -\pi i\delta(k_\theta^2)\epsilon(k_0), \ (k_\theta^2+i\epsilon)^{-2}\rightarrow \pi i\delta'(k^2_\theta)\epsilon(k_0)
\end{equation}
All the above results are compatible with the field equations.

Alternatively, the propagators may be derived from (36) as the vacuum expectation values of time-ordered products.  This confirms that the $+i\eta$ prescription for the double poles is consistent with (36).

The creation and annihilation operators used above
are at $t=0$. Using the free Hamiltonian $H_0$ in (45), we calculate their time dependence, giving
\begin{equation}
    a_-(\mathbf{k},t)=a_- e^{(-iK_\theta t)}, \ a_+(\mathbf{k},t)=(a_+ +isK_\theta t a_- )e^{(-iK_\theta t)}.
\end{equation}
These obey the same commutation relations as at $t=0$. (Up to this point, we have understood the creation and annihilation operators to  be at $t=0$.)

\section{The KO BRST operator}

The operator $Q$ which generates the increments (18) is
\begin{equation}
Q=\int d^3\mathbf{x}\{\Pi_i D_i c-(1/2)g\Pi_c . c\wedge c+\Pi_{\hat{c}}B\}.
\end{equation}
$Q$ is conserved, and commutes with $H$ and with the $S$ operator.

For asymptotic, free fields this is
\begin{equation}
    Q_0=\int d^3\mathbf{x}\{F_{oi}\partial_i c-\theta^2B\dot{c}\}.
\end{equation}
Integrating by parts and using (10), this gives
\begin{equation}
Q_0=\theta^2\int d^3\mathbf{x}\{c\dot{B}-B\dot{c}\}.
\end{equation}
In terms of annihilation and creation operators (from (39) and (41)),
\begin{equation}
    Q_0=\int d^3\mathbf{p} (ua_-^*+u^*a_-).
\end{equation}
$Q_0$ has the following commutation relations with annihilation operators:
\begin{equation}
    \{v,Q_0\}=a_-, \ \ [a_+,Q_0]=u.
\end{equation}

      After the transformation (30), the KO charge becomes
      \begin{equation}
 Q'=\int d^3\mathbf{x}[\Pi_iD_i c-g\Pi_c.c\wedge c/2-\theta^{-2} \Pi_{\hat{c}}.\Pi_0].        
      \end{equation}
The KO increments become, as well as the first two of (18),
      \begin{equation} \delta \Pi_i=\eta g\Pi_i \wedge c, \ \delta A_0=-\eta\theta^{-2}\Pi_{\hat{c}}, \ \delta\hat{c}=\eta\theta^{-2}\Pi_0, \
      \delta\Pi_c=\eta (D_i\Pi_i+g\Pi_c \wedge c).
      \end{equation}
      
\section{Generalisation and Lorentz covariance}
      In this section we set $\theta =1$, since otherwise we do not expect Lorentz covariance.
  The KO field $A$ in (36) assumes a preferred Lorentz frame  in two respects. It uses the basis vectors (32) and it mentions the time coordinate $t$.
  First, we replace the particular basis vectors $e$ in (32) by a more general basis, $e'$,  defined by the property that they can be obtained by the action on $e$  of Lorentz transformations that keep $p$ unchanged.  The $e'$ satisfy the following:
\begin{equation}
  e^{'2}_\pm=0,\ e'_+. e'_- =-1, \ e^{'m}_T.e_T^{'n}=\delta_{mn}, \ e^{'m}_T .e'_\pm=0,
  \end{equation}
  and
  \begin{equation}
      p.e'_+ =-\sqrt{2}P, \ p.e'_- =p.e^{'m}_T =0.
  \end{equation}

 A class of Lorentz transformations which
 preserve $p$ is
  \begin{equation}
e'_- =e_-,\ e_T^{'m}=e_T^m+\xi^m e_-,\ e'_ + =e_+-\sum\xi^m e_T^m -\frac{1}{2}e_-[(\xi^1)^2 +(\xi^2)^2],  \end{equation}
where $\xi^m$ (m=1,2) are two real numbers.
In addition there are rotations  between the two $e_T^m$, but for simplicity we omit them.

We will show that a general class of possible forms for the field $A_\mu$ has the structure
\begin{equation} 
\int E(a'_-  e'_{+\mu} +a'_+ e'_{-\mu}+\sum a_T^{'m} e_{T\mu}^{'m})+ (c.c.) +\partial_\mu\Lambda
\end{equation}
where
\begin{equation}
    \Lambda =\frac{s}{2}x^{\mu}\int  Ea_-[e'_{+\mu}  + ye'_{-\mu} +\sum(y^m e_{T\mu}^{'m})]+(h.c.)
\end{equation}
for any $y, y^m$ . The second term in (64) is a sort of gauge transformation.

The  field equations which (64) has to satisfy are (11) and (13), (14),(15). The latter three reduce for $\theta=0$ to the single covariant equation 
\begin{equation}
\partial^\nu \partial_\nu A_\mu=-s\partial_\mu B. 
\end{equation} 
Both (11) and (65) require the calculation of $\partial.\partial \Lambda$. The first term in (65) contributes $sa'_- $ to this, but the other terms contribute zero. This follows from (62).

The commutation  relations (25) and (26) are not affected by (65) because the only operators are $a_-$ and $a^*_-$, which commute.

It follows from (62) that if
\begin{equation}
    \Lambda'=\frac{s}{2}\int (Ea'_- x.e'_+  + h.c.)
\end{equation}
and
\begin{equation}
    \Lambda= \frac{s}{2}\int (Ea_- x.e_+ +h.c.),
\end{equation}
then (using $a_- =a'_-$)
\begin{equation}
    \Lambda'=\Lambda +\Omega
\end{equation}
where $\partial.\partial\Omega=0$; that is $\Omega$ is a (local) gauge transformation which does not alter $\partial.A$; and so the complete Lagrangian (5) is invariant under the  gauge transformation $\Omega$. Thus, under a change of the $e$ basis vectors, the second term in (64) is invariant up to a gauge transformation $\Omega$.
 
 In addition to the above, there is the operator
 \begin{equation}
     \omega =iy' s\int (a_- E  - a^*_- E^*)/P,
 \end{equation}
 which trivially satisfy $\partial_\mu \partial^\mu \omega =0$.
 
Under the transformation (63) of basis vectors, the first term in (64) undergoes a unitary transformation $U$, with 
\begin{equation}
U=\exp{\int d^3 k\sum{\xi^m (a_T^{m*} a_- -a^*_- a_T^m)}}.
\end{equation}
Under this, the annihilation operators transform as follows:
\begin{equation*}
a'_- =U^*a_-  U=a_-, \  a'_T =U^* a_T  U =a_T+\sum\xi^m a^m_-,
\end{equation*}
\begin{equation}
    a'_+ =U^* a_+ U=a_+- \sum \xi^m a_T^m -\frac{1}{2}a_- \sum \xi^m \xi^m.
    \end{equation}
These quantities have the property that
\begin{equation}
  a'_- e'_ + +a'_+ e'_ - + \sum a_T^{'m}e_T^{'m} = a_- e_+  +a_+ e_-+\sum a_T^{m}e_T^{m},
\end{equation}
which completes the proof of the invariance of the first line of ( 64).

Because the $s$ part of (64) is a gauge transformation, $F_{\mu\nu}$ has no $s$-dependence. Then, by (19) and (9),
neither do $\Pi_i$ or
$B$ (for free asymptotic fields).

The KO form equation (36) above, is recovered as a particular case of the parameters  $y, y^m, y'$ above, and replacing the vectors $e'$ by $e$:
\begin{equation*}
    y=-1,\ y^m =0, \ y'=1/(2\sqrt{2}).
\end{equation*}

\section{The transverse states}

We first recall the KO results. Let us write
\begin{equation}
    |F>, \ |A>, \ |T>
\end{equation}
for a general Fock state, an allowed state, and a pure transverse state (that is  a state containing transverse quanta only) respectively, where
\begin{equation}
   Q|A>=0.
    \end{equation}
    Then KO prove that
    \begin{equation}
     |A>=|T>+Q|F>.   
    \end{equation}
       Because the S-operator commutes with $Q$,
   \begin{equation}
       S|A>=|A'>=|T'>+Q|F'>
   \end{equation}    
and therefore
   \begin{equation}
       <A'|S|A> = <T'|S|T>.
   \end{equation}

These matrix elements should be independent of the Lorentz frame in which the transverse polarization vectors are defined. From the analysis in the previous section, we can see why this is so.
 The transformation (71) introduces states with one or more $a_T^*$ operators replaced by $a_-^*$ operators. These in turn can be replaced by $Qv^*$, and then, using that $Q$ commutes with the $S$ operator, we get zero.

\section{The structure of Fock space}
The restriction to allowed states raises the problem that many of these states have zero norm, meaning
that no  probability information  can be deduced about them. The exception is the class of pure transverse states, which have a positive norm; and, as shown above, S-matrix elements between these
states are all that is needed. However, if for example one wants to derive the imaginary part of an individual Feynman graph from unitarity, we need to include other Fock sates than the allowed ones.

That is the subject of this section, which is limited to the special case $s=0$.
Then the free Hamiltonian $H_0$ in (45)   is invariant
under the duality transformations
\begin{equation}
  a_\pm \rightarrow \tilde{a}_\pm =a_\mp, \ u\rightarrow \tilde{u}=v, \ v\rightarrow \tilde{v }=u.
\end{equation}
This gives the dual of $Q_0$
\begin{equation}
    \tilde{Q}_0=\int_\theta  d^3 k (va_+^* +v^*a_+),
\end{equation}
which commutes with $H_0$. (but not of course with the complete $H$). Also
\begin{equation}
    \{Q_0,\tilde{Q}_0\}=N,
\end{equation}
where
\begin{equation}
    N=\int_\theta (a_+^*a_- +a_-^* a_+ +u^*v+v^*u)
\end{equation}
is the operator which counts the number  of creation operators other then the transverse ones.

Let $P_n$ be the projection operator which projects onto
Fock states $|>$ which contain $n$ of the operators $a_\pm^*, u^*, v^*$ (and any number of transverse operators).
This may be expressed as
\begin{equation}
    P_n=\frac{1}{\pi}\int_0^\pi d\phi\cos[(N-n)\phi]
\end{equation}
which is Hermitian.

From (81), for any state $|F>$,
\ \begin{equation}
    |F>=|F_Q>+|F_{\tilde{Q}}>+P_0|F>
\end{equation}
where
\begin{equation}
    |F_Q>=\sum_{n>0}Q_0\tilde{Q}_0(P_n/n)|F>
\end{equation}
and
\begin{equation}
    |F_{\tilde{Q}}>=\sum_{n>0}\tilde{Q}_0 Q_0(P_n/n)|F>.
\end{equation}
Then
\begin{equation}
    Q_0|F_Q>=0, \ \tilde{Q}_0|F_{\tilde{Q}}>=0,
\end{equation}
and
\begin{equation}
    <F_Q|F_Q>=0, \ <F_{\tilde{Q}}|F_{\tilde{Q}}>=0.
\end{equation}
Then it follows that
\begin{equation}
    <F|F>=<F_Q|F_{\tilde{Q}}>+<F_{\tilde{Q}}|F_Q>+<F|P_0|F>.
\end{equation}
It follows that exactly half  of the Fock states , other than the pure transverse states, are allowed states.
\bibliographystyle{plain}

\end{document}